\documentclass{jpsj3}
\usepackage{txfonts}
\usepackage{graphicx}
\usepackage{color}
\usepackage{lineno}

\title{First-principles study on structure and anisotropy of high N-atom density layer in 4H-SiC}

\author{Mitsuharu Uemoto$^1$\thanks{uemoto@eedept.kobe-u.ac.jp}, Naoki Komatsu$^1$, Yoshiyuki Egami$^2$ and Tomoya Ono$^1$}
\inst{$^1$Department of Electrical and Electronic Engineering, Graduate School of Engineering, Kobe University, 1-1 Rokkodai-cho, Nada-ku, Kobe 651-8501, Japan \\
$^2$Division of Applied Physics, Faculty of Engineering, Hokkaido University, Sapporo, Hokkaido 060-8628, Japan} 

\abst{
A nitridation annealing process is well employed to reduce interface trap states that degrade the channel mobility of 4H-SiC/SiO${}_2$ metal-oxide-semiconductor field-effect transistor. In recent experiments, the existence of high N-atom density layers at the annealed interface is reported and their concentrations are known to be anisotropic in the crystal planes. Until now, the role of atomic structure and the electronic states surrounding the N atoms incorporated by the nitridation annealing process on the origin of anisotropy is not well understood. In this work, we propose a simplified atomic-scale model structure of 4H-SiC with the a high N-atom density layer ($\sim 10^{15}~\mathrm{atom}/\mathrm{cm}^2$), which is of the order of the experimental observation. We use bulk 4H-SiC as host crystal and consider several sets of the atomic configurations of the N-atom incorporated structure at the quasi cubic-($k$-) and hexagonal-($h$-)sites on $a$-, $m$-, and Si-(C-)planes. Based on the density functional theory calculations, we investigate the influence of the energy stability on the distribution directions. Although our bulk model is simplified compared to the realistic interface structures, we confirm significant difference among models and observe that the incorporation of N atoms on the $a$-face is stable. Furthermore, from the analysis of the electronic states, we suggest that this anisotropy of the formation energy originates from the change of the coordinating number due to the difference in geometric configurations of the N-atom incorporated structures.
}

\begin{document}

\maketitle

\section{\label{sec:intro} Introduction}

4H-silicon carbide (4H-SiC) is an IV-IV type semiconductor with a wide band gap of $3.2~\mathrm{eV}$~\cite{levinshtein2001properties}, and the 4H-SiC based metal-oxide-semiconductor field-effect transistor (MOSFET) is expected to be used in next generation switching devices operating at high power and high-frequency applications \cite{hijikata2012physics, kimoto2015material}. To reduce the density of interface defects, a passivation treatment with nitrogen has been used and improvement in the mobilities $\mu = 30 \sim  80~\mathrm{cm}/\mathrm{V}\cdot\mathrm{s}$ has been achieved \cite{fiorenza2019characterization, masuda2019demonstration}. The experimental characterizations of the annealed SiC/SiO${}_2$ interface are previously reported \cite{hamada2017analysis,kosugi2011fixed}, where the incorporated N atoms are fixed on the SiC substrate side at the C-atom sites that are chemically bonded with the Si atoms. Hamada \textit{et al.} measured  N-atom density by the secondary ion mass spectrometry and reported the presence of a high N-atom density layer.\cite{hamada2017analysis} The 4H-SiC has several crystal planes; using the conventional expressions of the hexagonal lattice, the faces can be described as the $a$-face $(1 \overline{1} 0 0)$, $m$-face $(1 1 \overline{2} 0)$, Si-face $(0 0 0 1)$, and C-face $(0 0 0 \overline{1})$.
For each crystal plane, the N-atom density reaches the order of $10^{14} \sim 10^{15}~\mathrm{atoms}/\mathrm{cm}^2$\cite{hamada2017analysis}; we can say that the most of C atoms at the interface are substituted by N atoms. Shirasawa \textit{et al.} proposed a model of N-annealed 6H-SiC/SiO${}_2$ interface, which constitutes epitaxially-stacked SiO${}_2$ and Si${}_3$N${}_2$ monolayers (SiON) on the SiC \cite{shirasawa2007epitaxial}. In the case of 4H-SiC, a similar structure was proposed and the simulated X-ray absorption spectroscopy spectrum agreed well with the experiment \cite{isomura2019distinguishing}.

Besides, there are some experiments reporting that this annealing process exhibits an anisotropic behavior with respect to the crystal planes \cite{dhar2005nitridation,hamada2017analysis}. Dhar \textit{et al}. measured the 4H-SiC/SiO${}_2$  interface by ${}^{15}$N nuclear reaction analysis and reported that the concentration of N atoms on the $a$-face is more than twice as compared to that on the Si-face \cite{dhar2005nitridation}. Reference \citen{hamada2017analysis} indicates that the $m$-face has nearly the same concentration of N atoms as the $a$-face; regardless of the annealing temperature, the N incorporation occurs in the order of $a \sim m \gg \mathrm{Si}$, excluding C-face. The anisotropy mechanism is yet to be well understood and the atomic-scale structure has not been understood well for arbitrary crystal planes.

In this work, we propose a universal atomic-scale model describing 4H-SiC with high N-atom density layer in arbitrary crystal planes. Besides, to understand the anisotropy of N-atom density, we study the stability and electronic states of the 4H-SiC, including the various configurations of the Si vacancy ($\mathrm{V}_\mathrm{Si}$) and N atoms at the C-atom sites ($\mathrm{N}_\mathrm{C}$). The areal density of N atoms in our model is $\sim 10^{15}~\mathrm{atoms}/\mathrm{cm}^2$, which agrees well with experimental results. Our structures do not generate any defect states in the band gap of 4H-SiC and the N-atom incorporation is thermodynamically preferable. From the structural optimization, we find the difference in formation energy for the N-atom incorporated structures among crystal planes, which is caused by the geometry of the chemical bonds around N atom in 4H-SiC; our results show that the incorporation at the $k$-site on the $a$-face is energetically stable compared to other sites and crystal planes. The wave functions of the conduction band minimum (CBM) are separated by the atomic planes containing N atoms. We predict that the \color{red} conducting properties \color{black} degrades in the direction perpendicular to the plane containing the N atoms. On the other hand, the order of formation energies in our bulk model is slightly different from that of the N-atom density in the experiments,\cite{dhar2005nitridation, hamada2017analysis} suggesting that the anisotropy of nitridation annealing process observed in the present experiments is due to the effect from the surface of the crystal plane and the kinetic process during the chemical reaction.

This paper is organized as follows: in Sec.~\ref{sec:method}, we present the crystal structures considered in this study and the computational conditions.
In Sec.~\ref{sec:result}, we report the optimized energies of each structure and discuss the electronic structures in terms of the density of states (DoS) and the wave function at the CBM.
Finally, in Sec.~\ref{sec:summary}, we conclude our study.

\section{\label{sec:method} Method}
\begin{figure*}
\includegraphics[width=0.90\textwidth]{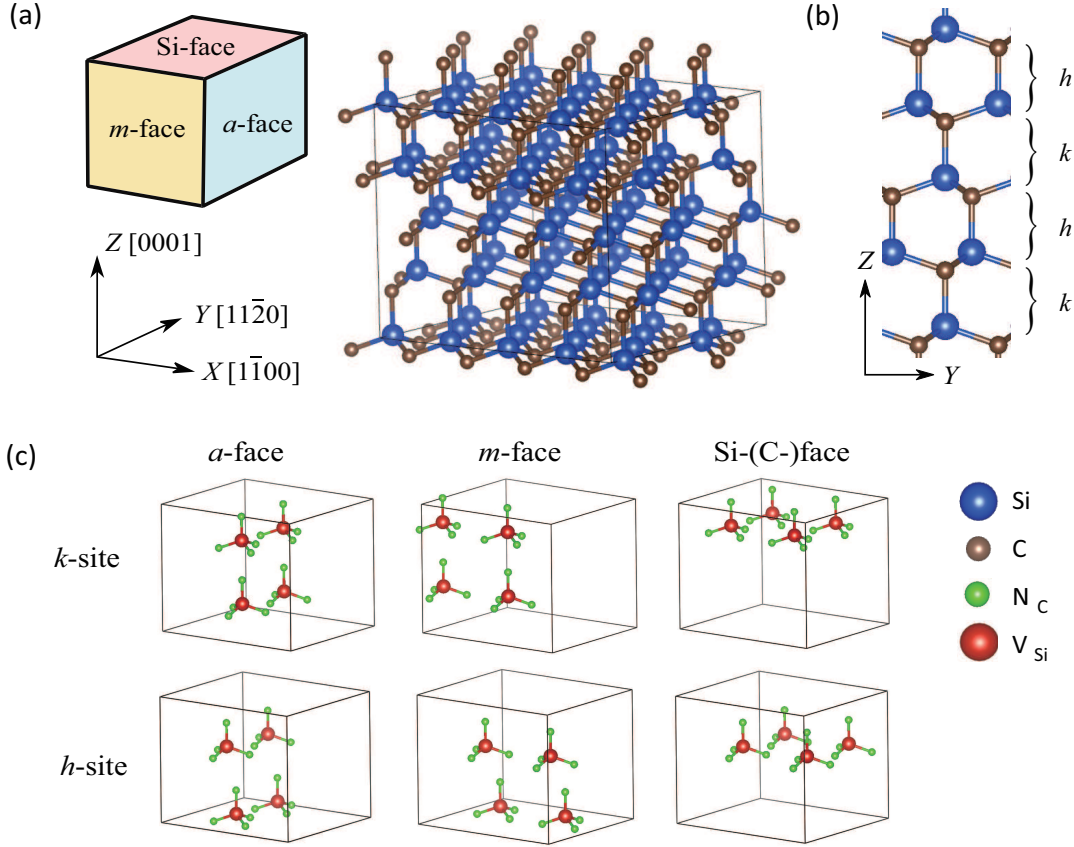}
\caption{
    \label{fig:crystal}
    Schematic of N-atom incorporated structures.
    (a) 4H-SiC supercell structure containing 128 atoms.
    (b) $k$- and $h$-site positions from $X [1\overline{1}00]$-direction.
    (c) Configuration of N-atom incorporated structures in host structure for 6 $L$-type arrangements.
}
\end{figure*}

In our model, for the host material of 4H-SiC, we employ a rectangular supercell of dimensions $12.3~\mathrm{\AA} \times 10.7~\mathrm{\AA} \times 10.1~\mathrm{\AA}$ containing 128 atoms \cite{levinshtein2001properties} [See. Fig.~\ref{fig:crystal}(a)]. 
For convenience, we refer to the three axes as $X~([1 \overline{1} 0 0])$, $Y~([1 1 \overline{2} 0]$, and $Z~([ 0 0 0 1 ])$; the cross sectional planes of $XY$, $YZ$, and $ZX$ represent the Si-(C-), $a$-, and $m$-faces, respectively.

We consider several modifications incorporating 4 $\mathrm{V}_\mathrm{Si}$s and 16 $\mathrm{N}_\mathrm{C}$s. Both Si and C atoms have four valence electrons, and a N atom has three valence electrons. We expect that a stable structure will not contain dangling bonds, from the similarity of the SiON layer model reported by Shirasawa \textit{et al}. \cite{shirasawa2007epitaxial,shirasawa2009atomic} 
To evaluate the anisotropy, we consider the cases in which $\mathrm{V}_\mathrm{Si}$s are arranged on the Si-(C-), $m$-, and $a$-faces. The model of the host structure is shown in Fig.~\ref{fig:crystal}(a). The areal density of N atoms in each crystal plane is shown in Table~\ref{tbl:density}; the proposed model can provide the layer with high N-atom density that reaches the order of $10^{15}~\mathrm{atom}/\mathrm{cm}^2$ reported in previous studies \cite{hamada2017analysis}.
In addition, the 4H-SiC has two inequivalent lattice sites, \textit{i.e.} $h$- (hexagonal-) and $k$- (quasi-cubic-) sites [see Fig.~\ref{fig:crystal}(b)]. With the possible options of the crystal planes [$a$, $m$, or Si(C)] and the lattice sites ($k$ or $h$), there are two possible methods to place $\mathrm{V}_\mathrm{Si}$ in a plane: In $L$ (linear) type geometry, $\mathrm{V}_\mathrm{Si}$s are aligned with each other, and in $S$ (staggered) type geometry, $\mathrm{V}_\mathrm{Si}$s are placed out of alignment with each other.
There are $3 \times 2 \times 2 = 12$ non-equivalent atomic configurations.

The first-principles calculations are performed by the RSPACE code \cite{rspace2005}, which uses the real-space finite-difference approach within the frameworks of density functional theory \cite{chelikowsky1994finite}. The local density approximation (LDA) is used for the exchange-correlation energy.
We select the Vosko-Wilk-Nusair parameterization of the LDA \cite{vosko1980accurate} and the projector-augmented wave (PAW) method is used for electron-ion interactions \cite{blochl1994projector}. A $3 \times 3 \times 3$ Monkhorst-pack $k$-point mesh including a $\Gamma$-point in the Brillouin zone (BZ) is adopted. The real-space grid in the supercell is $60 \times 50 \times 48$, which has a spacing of about $0.21~\AA$. We perform geometry optimization until the atoms' residual forces are smaller than 0.001 Hartree/Bohr radius.

\begin{table}[tbp]
\caption{
N-atom density for each crystal plane in Fig.~\ref{fig:crystal}(c).
\label{tbl:density}
}
\begin{tabular}{ccc}
\hline
 Face & N-atom density~[$\mathrm{atom}/\mathrm{cm}^2$] \\ \hline
$a$ & $1.48 \times 10^{15}$ \\
$m$ & $1.29 \times 10^{15}$ \\
$\mathrm{Si(C)}$ & $1.22 \times 10^{15}$ \\
\hline
\end{tabular}
\end{table}

\section{\label{sec:result} Results and Discussion}
\begin{table}[tbp]
\caption{N-atom coordinating numbers and band gaps $E_\mathrm{gap}$ for the considered modifications of N-atom incorporated structures. Each modification is characterized by the crystal plane [$a$, $m$, and Si(C)] and the lattice site (hexagonal $h$ or quasi-cubic $k$). $\mathrm{Si}(0)$ represents those number of Si atoms that are not neighboring the $\mathrm{N}_\mathrm{C}$ atoms; whereas $\mathrm{Si}(1)$ and $\mathrm{Si}(2)$ represent those neighboring one and two $\mathrm{N}_\mathrm{C}$ atoms, respectively.
}
\label{tbl:energy}

\begin{tabular}{ccccccc}
\hline
        & Site & Face & Si$(0)$ & Si$(1)$ & Si$(2)$ &$E_\mathrm{gap}$~(eV) \\ \hline
        & $k$  & Si(C)& 24      & 24      & 12      & 2.223                \\
        & $k$  & $m$  & 28      & 16      & 16      & 2.413                \\
        & $k$  & $a$  & 32      & 8       & 20      & 2.455                \\ \cline{2-7} 
        & $h$  & Si(C)& 24      & 24      & 12      & 2.120                \\
        & $h$  & $m$  & 24      & 24      & 12      & 2.212                \\
        & $h$  & $a$  & 28      & 16      & 16      & 2.376                \\ \hline
\end{tabular}
\end{table}

\begin{figure}
\includegraphics[width=0.5\textwidth]{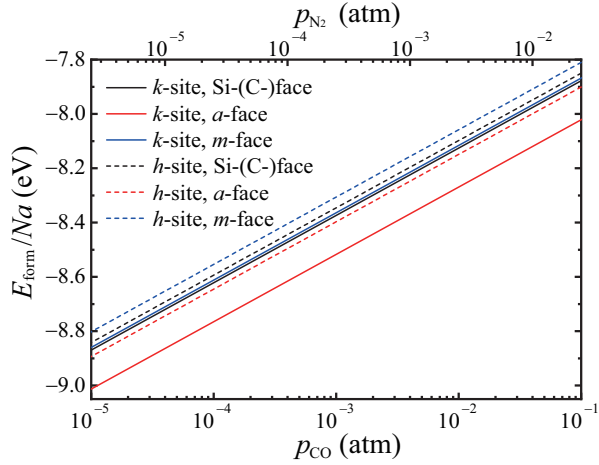}
 \caption{
     \label{fig:formationenergy}
     Formation energies $E_\mathrm{form}$ with respect to partial pressures of CO and N$_2$. $Na(=4)$ represents the number of N$_\mathrm{C}$ per one V$_\mathrm{Si}$.
}
\end{figure}

The optimized lattice constants of SiC with the N-atom incorporated structures are $a$=2.98 \AA \hspace{2mm} and $c$=4.87 \AA, which are comparable with that of 2H-SiC ($a$=3.07 \AA \hspace{2mm} and $c$=4.92 \AA), indicating that the dangling bonds are hardly generated by the lattice constant mismatch at the interface. For the modification of $\mathrm{V}_\mathrm{Si}$ and $\mathrm{N}_\mathrm{C}$ positions, we determine the following formation energy.
\color{red}
\begin{align*}
E_\mathrm{form}=& 
E(\mathrm{SiO}_2) 
+ \frac{1}{4}E_{total}^{(SC)}(4 \mathrm{V}_\mathrm{Si} 16 \mathrm{N}_\mathrm{C})
+ 4\mu_\mathrm{CO} 
+ 2\mu_\mathrm{N} 
- 16 E(\mathrm{SiC}) 
- 6 \mu_\mathrm{NO},
\end{align*}
where $E(\mathrm{SiC})$ and $E(\mathrm{SiO}_2)$ are total energies of SiC, $\mathrm{SiO}_2$ units in bulks of 4H-SiC and quartz SiO$_2$, respectively. 
$\frac{1}{4} E_\mathrm{total}^\mathrm{(SC)}(4 \mathrm{V}_\mathrm{Si} 16 \mathrm{N}_\mathrm{C})$ represents the total energies of considered modifications with $4 \mathrm{V}_\mathrm{Si}$ and $16 \mathrm{N}_\mathrm{C}$ in the 4H-SiC supercell.
\color{black}
In addition, $\mu_\mathrm{NO}$, $\mu_\mathrm{CO}$, and $\mu_\mathrm{N}$ are the chemical potentials of NO molecules, CO molecules, and N atoms in N$_2$ molecules, respectively. The temperature is set at 1000 K and the partial pressure of NO gas, $p_\mathrm{NO}$, is 1 atm. Since the annealing process is not equilibrium states, it is not straightforward to determine the partial pressures of CO, $p_\mathrm{CO}$, and N$_2$, $p_{\mathrm{N}_2}$, gases. Here, $p_\mathrm{CO}$ ($p_{\mathrm{N}_2}$) is varied between $10^{-1}$ atm and $10^{-5}$ atm ($0.25 \times 10^{-1}$ atm and $0.25 \times 10^{-5}$ atm). The formation energies with respect to $p_\mathrm{CO}$ and $p_{\mathrm{N}_2}$ are shown in Fig.~ \ref{fig:formationenergy}. Here, we concentrate on the 6 $L$-type modifications, which provide energetically stable results [see Fig.~\ref{fig:crystal}(c)], since the $S$-type arrangements are more unstable than the $L$-type arrangements. According to Fig.~ \ref{fig:formationenergy}, the following trends are observed: (1) all models considered here are stable because the formation energies are negative. (2) Incorporation at the $k$-site is more stable than $h$-site for the same crystal plane. (3) Incorporation on the $a$-face is always more stable than the other faces for the same sites. Although the areal N-atom density in the present models is comparable with those observed in the experiments, the order of the formation energies is not consistent with that of the areal N-atom densities in the experimental results, implying that the effects from the surface of crystal planes and the reaction process limit the areal N-atom density in the present experiments. Indeed, the experimental study reported the large difference in the N-atom density between the Si-face and C-face \cite{hamada2017analysis}. However, our bulk model does not distinguish between Si-$(0 0 0 1)$ and C-$(0 0 0 \overline{1})$-planes. Our results also indicate that the atomically flat interface can be fabricated on $a$-plane by optimizing the annealing process.

We suggest that this anisotropy of the formation energy is related to the number of Si-N bonds around the substitutional sites. In Table~\ref{tbl:energy}, the first three columns represent the number of Si atoms neighboring 0, 1, and 2 $\mathrm{N}_\mathrm{C}$ atoms; $\mathrm{Si}(0)$ represents the number of Si atoms that are not neighboring $\mathrm{N}_\mathrm{C}$ atoms (connected to four carbon atoms), and $\mathrm{Si}(1)$ and $\mathrm{Si}(2)$ are those neighboring one and two $\mathrm{N}_\mathrm{C}$ atoms, respectively. The energetically most stable configuration ($k$-site and $a$-face) has the largest number of $\mathrm{Si}(2)$.

\begin{figure}
\includegraphics[width=0.5\textwidth]{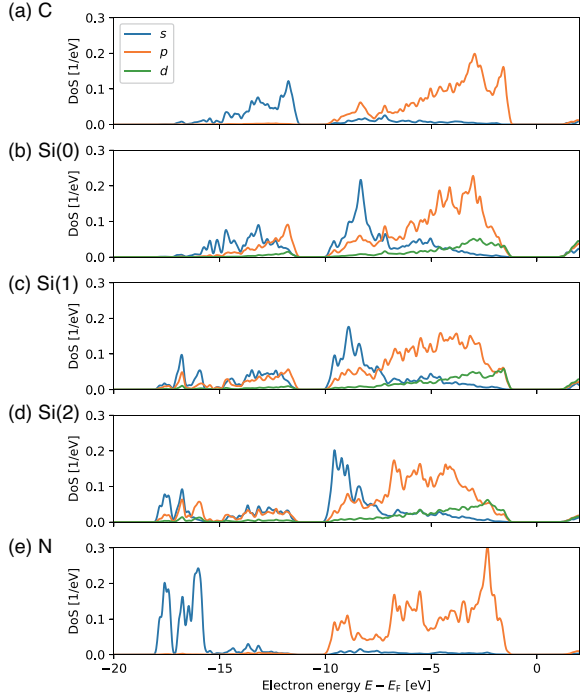}
 \caption{
     \label{fig:pdos}
     PDoS of various sites in a modification (incorporation at $k$-site and $a$-face incorporation).
     (a) PDoS at a C atom distant from $\mathrm{N}_\mathrm{C}$.
     (b) PDoS at a Si atom distant from $\mathrm{N}_\mathrm{C}$.
     (c) PDoS at a Si atom neighboring one $\mathrm{N}_\mathrm{C}$.
     (d) PDoS at a Si atom neighboring two $\mathrm{N}_\mathrm{C}$s.
     (e) PDoS at a $\mathrm{N}_\mathrm{C}$.
}
\end{figure}
To understand the effect of the neighboring N atoms on the electronic structures, we calculate the partial density of states (PDoS). The PDoS is evaluated from the expansion coefficient of the atom of interest, obtained by expanding the wave function into pseudo partial waves in the PAW method. The PDoS profiles are smoothened by a Gaussian broadening function with a full-width half-maximum (FWHM) of $0.16~\mathrm{eV}$. 

Figures~\ref{fig:pdos}(a) and (b) represent the PDoS at the $\mathrm{C}$ and $\mathrm{Si}$ sites, located away from N atoms. They are expected to be nearly the same in the bulk 4H-SiC; the site in Fig.~\ref{fig:pdos}(b) is counted as $\mathrm{Si}(0)$ in Table~\ref{tbl:energy}. In the bulk 4H-SiC, all of the four $sp^3$ orbitals of Si atoms hybridize with $sp^3$ orbitals of neighboring four C atoms and form the valence bands. As can be seen in Figs.~\ref{fig:pdos}(c) and (d) (counted as $\mathrm{Si}(1)$ and $\mathrm{Si}(2)$ in Table~\ref{tbl:energy}, respectively), the peak positions of the bands shift to lower energy from the case of (b). In such cases, $sp^3$ orbitals of Si atoms hybridize with $p$ orbital of N atoms instead of C atoms. N atoms have a larger nuclear charge than C atoms.
Therefore, the electron energy is expected to decrease when the number of Si-N bonds increases.

\begin{figure}
\includegraphics[width=0.60\textwidth]{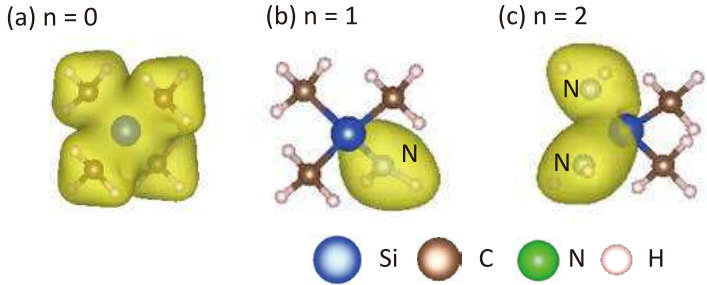}
 \caption{
     \label{fig:orbit}
     Orbital charge profile of the lowest energy bonding orbitals of Si(NH${}_2$)${}_n$ (CH${}_3$)${}_{4-n}$.
}
\end{figure}

\begin{table}[htbp]
\caption{
\color{red}
Total energies $E_n$ for $\mathrm{Si}(\mathrm{NH}_2)_{n}(\mathrm{CH}_3)_{4-n}$ molecules and energy difference $\Delta{E}_n$ defined as Eq.~(\ref{eq:model_delta}).
The constant term $E(\mathrm{C}) + E(\mathrm{H}) - E(\mathrm{N})$ is $119.006$~[eV].
\color{black}
}
\label{tbl:model}
\begin{tabular}{llll}
\hline  \hline
$n$ & Compond        & $E_n$~[eV]   & $ \Delta E_n$~[eV]      \\
\hline 
0 & Si(CH${}_3$)${}_4$       & $-1767.216$ &            \\
1 & Si(NH${}_2$)(CH${}_3$)${}_3$  & $-1883.064$ & $3.158$ \\
2 & Si(NH${}_2$)${}_2$(CH${}_3$)${}_2$ & $-1999.014$ & $3.056$ \\
3 & Si(NH${}_2$)${}_3$(CH${}_3$)  & $-2114.956$ & $3.063$ \\
4 & Si(NH${}_2$)${}_4$       & $-2230.921$ & $3.041$ \\
\hline  \hline 
\end{tabular}
\end{table}

Moreover, we also consider a simplified model, molecules of $\mathrm{Si}(\mathrm{NH}_2)_{n}(\mathrm{CH}_3)_{4-n}$, containing $n$ Si-N bonds. As seen in Table~\ref{tbl:model}, we compute the total energy $E_n$ and $\Delta{E}_n$ using the following equation:
\color{red}
\begin{equation}
    \Delta E_n
    =
    E_n
    -
    E_{n-1}
    +
    \left[
        E(\mathrm{C}) + E(\mathrm{H}) - E(\mathrm{N})
    \right]
    \label{eq:model_delta}
    \;,
\end{equation}
which represents the energy change caused by adding one more neighboring N atom. 
$E(\mathrm{C})$, $E(\mathrm{H})$ and $E(\mathrm{N})$ are the total energies of single C, H and N atoms, respectively. 
\color{black}
In Table~\ref{tbl:model}, we find the relation of $\Delta{E}_2 < \Delta{E}_1$; the formation of Si atoms with two $\mathrm{N}_\mathrm{C}$ atoms is energetically more stable than that with one $\mathrm{N}_\mathrm{C}$ atom.

The spatial profiles of the wave functions are useful for understanding the relation of $\Delta E_n$. In the case of $n=0$, the Si atom is surrounded by four equivalent CH${}_3$ branches; there are four degenerated $sp^3$ orbitals of a Si atom. The lowest bonding state represents the hybridization of these four $sp^3$ orbital [see Fig.~\ref{fig:orbit}(a)]. In the case of $n=1$, the Coulomb potential of the N atom perturbes one of the $sp^3$ orbitals and the lowest bonding state consists of the $sp^3$ orbital directing to the N atom [see Fig.~\ref{fig:orbit}(b)]. The localization of this wave function around the N atom leads the increase of its kinetic energy. In the case of $n=2$, there are the two $sp^3$ orbitals on N atoms, and the wave function of the lowest bound state is distributed around two N atoms [see Fig.~\ref{fig:orbit}(c)]. Since the wave function is not as localized as for $n=1$ case, and the kinetic energy by the confinement is decreased in the case $n=2$. We consider the localization of the wave function to be the origin of the relative stability of the $n=2$ configuration.

The above results indicate that the presence of Si atoms connected to two $\mathrm{N}_\mathrm{C}$ atoms can reduce the total energy. This is consistent with the relationship between the $\mathrm{N}_\mathrm{C}$ coordinating number and formation energy seen in Table~\ref{tbl:energy} and Fig.~\ref{fig:formationenergy}, respectively. Therefore, we suggest that the anisotropy of the formation energy originates from the coordinating number due to the difference in geometric configurations of the N-atom incorporations. As seen in Table~\ref{tbl:energy}, the above mechanism results in the following tendencies, where incorporation of N atoms on the $a$-face is more stable than the Si-(C-) and $m$-faces, and the $k$-site is more stable than the $h$-site.

\begin{figure*}
\includegraphics[width=0.8\textwidth]{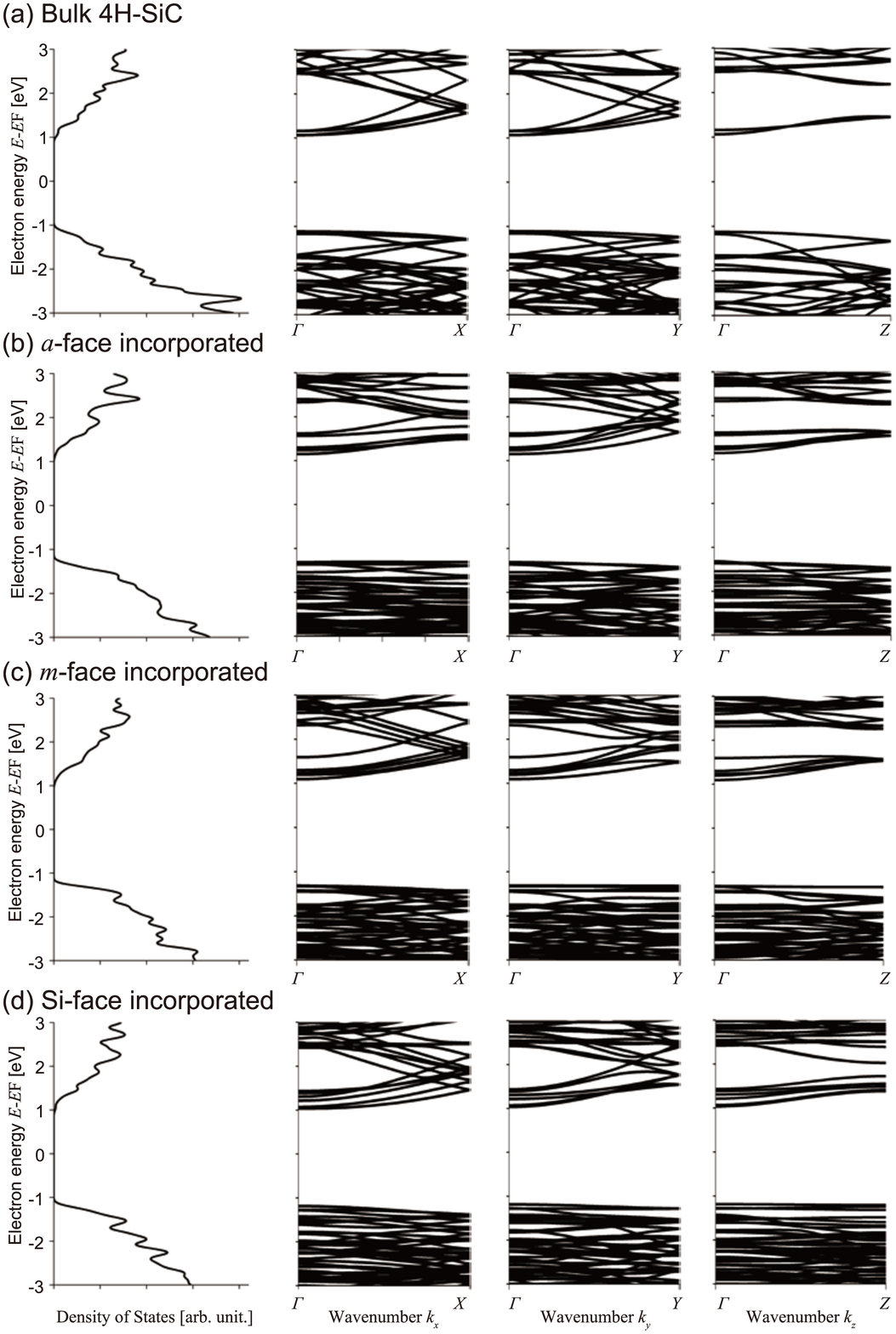}
 \caption{
     \label{fig:band}
     Electronic structures of (a) pristine and (b)-(d) SiC with N-atom incorporated structures on Si-(C-), $m$-, and $a$-faces, respectively.
     The first column shows the total DoS; the second, third, and forth columns are the band dispersion along the $k_z$($\Gamma$-$Z$), $k_x$($\Gamma$-$X$), and $k_y$($\Gamma$-$Y$) directions, respectively.
     The Fermi level defines the zero of the energy (vertical direction).
     We employ the same FWHM as that in Fig.~\ref{fig:pdos}.
}
\end{figure*}

Next, we focus on the band gap of each modification represented by $E_\mathrm{gap}$ in Table~\ref{tbl:energy}. For the pristine 4H-SiC, $E_\mathrm{gap}=2.138~\mathrm{eV}$, which is slightly smaller than the experimentally observed value ($E_\mathrm{gap}=3.26~\mathrm{eV}$) \cite{kimoto2015material}, due to the well-known underestimation behavior by the LDA \cite{vosko1980accurate}. With the incorporation of N atoms, the band gap tends to increase. The increase is significant for the $a$- and $m$-faces, while the change is relatively small for the Si-(C-)face. In particular, the DoS and band structures are shown in Fig.~\ref{fig:band} for the incorporation at the $k$-site, which has highly stable structures.
\color{red}
In addition, there are no in-gap state in the calculated electronic band structure; it indicates the absence of dangling bond in our model.
\color{black}

The band structure of the 4H-SiC primitive cell has a CBM at the $M$-point that exists on the $\Gamma-X$ ($[1 \overline{1} 0 0]$-direction). In the case of our rectangular supercell, the $M$-point is folded to the $\Gamma$-point of the reduced BZ. As seen from Fig.~\ref{fig:band}(a), the CBM is observed at $\Gamma$-point. The band dispersion around the CBM is anisotropic; there is a flat and isolated band along the $Z$-($[0 0 0 1]$) direction. This is due to the low periodic symmetry of stacking along the $Z$-direction in 4H-SiC. The effective mass tensor increases nearly twice as much in the $Z$ direction compared to other directions \cite{lambrecht1995band}; consequently, its conduction properties degrade.

\begin{figure}
\includegraphics[width=0.5\textwidth]{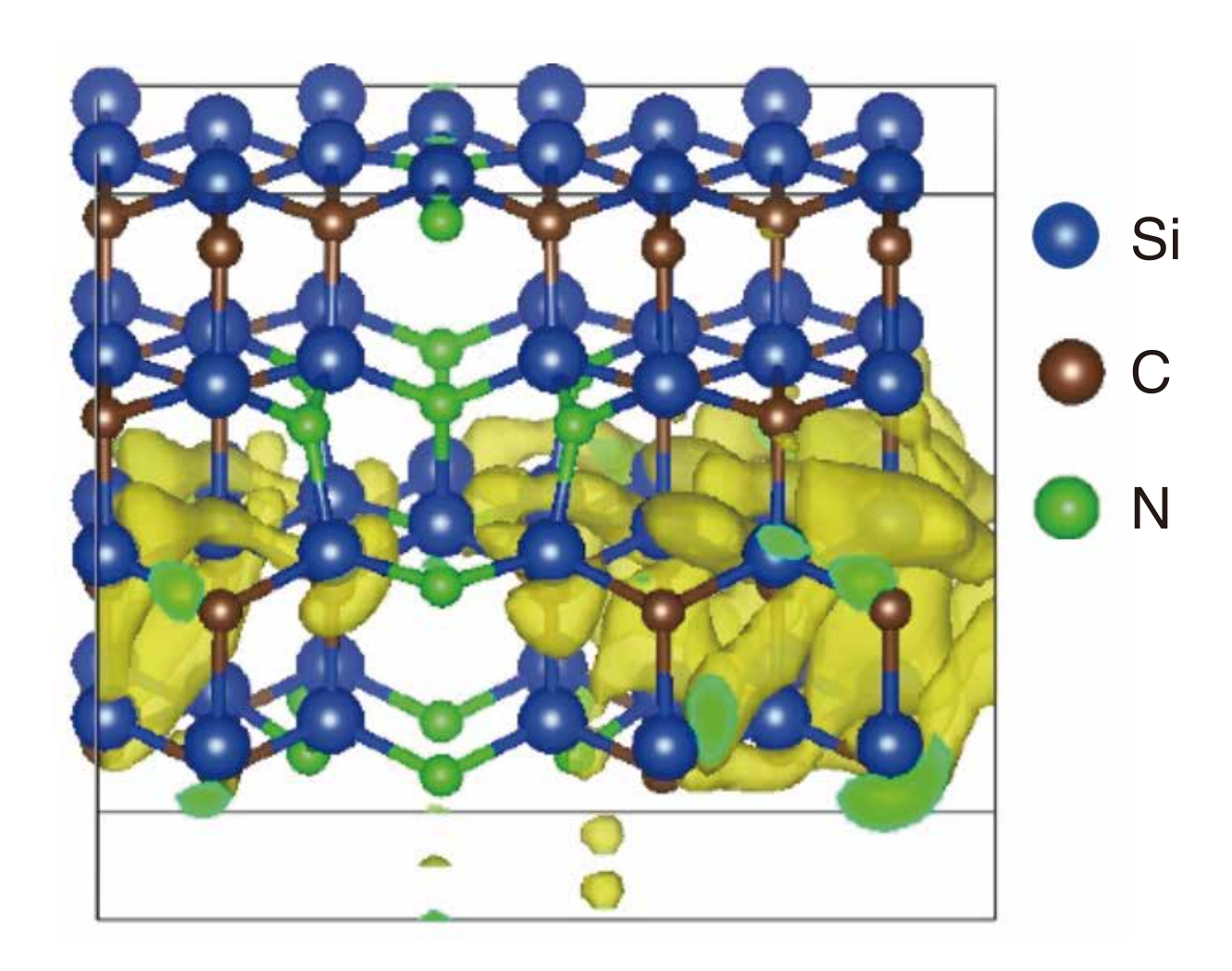}
 \caption{
     \label{fig:wavefunc}
    Charge density distribution of wave function of CBM in 4H-SiC with incorporation at $k$-site and $a$-face.
}
\end{figure}

Figure~\ref{fig:band}(b) represents the cases for the N-atom incorporation on the $a$-face. The modulation of the conduction band arises in the $X$ ($[1 \overline{1} 0 0]$)-direction, and we can observe a gap in the degeneracy around the $X$-point. These behaviors are due to the symmetry breaking caused by the wall-like N-atom incorporated structures. The conductivity decreases in the $X$-direction. When the N-atom incorporated structures are distributed on $ZX$ or $YZ$ plane, the quantum confinement effect increases the energy levels of the CBM. This is why the band gap increases in the case of incorporation on the $a$- and $m$-faces (see Table~\ref{tbl:energy}). In the 4H-SiC, the CBM has a characteristic wave function, namely the floating state \cite{matsushita2012floating}, which is not distributed in the vicinity of the atomic sites; the high N-atom density layer breaks this wave function (see Fig.~\ref{fig:wavefunc}). Our earlier study noted that such incorporations at the interface may reduce the carrier mobility of the MOSFET due to the scattering of the conduction electrons \cite{ono2017theoretical}. For example, in the case of nitridation annealing on the Si-(C-)face, the nitridation is expected to grow on the $a$-face due to the low formation energy when the effect of the crystal surface is ignored. 
The wall-like incorporations, which break the wave function of the CBM, grow the band gap and decrease the carrier mobility in the $X$-direction $(1 \overline{1} 0 0)$, are introduced in the nitridation annealing on the $m$- or Si-(C-)face.
Therefore, the $m$- or Si-(C-)face based MOSFETs are undesirable due to the degradation in the conduction properties. We believe that the nitridation annealing of the $a$-face is preferable for the 4H-SiC based MOSFET.

\section{\label{sec:summary} Summary}
We conducted a first-principles study on atomic-scale mechanisms leading to the anisotropy of the areal N-atom density in the nitridation annealing process of 4H-SiC based MOS. We proposed the simplified model of bulk 4H-SiC with the high N-atom density layers, considered various atomic configurations of the N-atom incorporated structures distributed across a plane, and analyzed their energetic stability and the influence of the crystal planes and sites. It is found that the proposed structures are enegetically stable, the lattice constant of the proposed structure is in accordance with that of SiC, and no defect states are generated in the band gap of SiC. As seen in Table~\ref{tbl:density}, the proposed model can provide the high N-atom density layer in which the density is similar order with that reported in the experiments ($\sim 10^{15}~\mathrm{atoms}/\mathrm{cm}^2$).\cite{dhar2005nitridation, hamada2017analysis} In addition, we showed that the distribution on the $a$-face at the $k$-sites is more stable compared to other faces and sites. The analysis of the electronic structure provides the mechanism for the anisotropy of the formation energy. Our results indicate that the number of Si atoms neighboring two $\mathrm{N}_\mathrm{C}$ atoms depends on the orientation, which is responsible for the anisotropy. 
The calculated order of the formation energy anisotropy is not consistent with the areal N-atom densities in the experimental results $(\mathrm{C} > a \sim m > \mathrm{Si})$.
\color{red}
We expect that this discrepancy is explained by two reasons. First, our bulk model does not describe the realistic SiC/SiO${}_2$ interface. In the 4H-SiC based MOSFETs, the thermal oxidation and nitridation annealing processes on the 4H-SiC surface create the N-atom incorporated structures on the substrate side of the SiC/SiO${}_2$ interface, and the configuration of N atoms is also expected to be affected by the neighboring O atoms. To obtain more realistic predictions, an improved understanding of the N-atom incorporation on various orientations of the MOS interface is desired since the effect from the surface of crystal plane is difficult to treat in the bulk model. Second, since we investigate the preferential direction of the growth of N-atom incorporated structures, our computational model does not include a kinetic process in which a N atom on SiO${}_2$ surfaces overcomes reaction barrier and intrudes into the SiC layer. Although our formation energy calculation indicates that the a plane is suitable for fabrication of atomically flat MOS interface, the kinetic process of N atom diffusion through SiO${}_2$, which should be suppressed in the future, plays important role in the present annealing process.
\color{black}

Moreover, a similar atomic structure is desired in the experimental and theoretical works for the NV-center applications in the quantum information \cite{csore2017characterization}. We expect a similar ``coordinating number''-based understanding of the distribution of the N-atom incorporated structures in crystals to our approach.

\begin{acknowledgments}
This work was partially supported by the financial support from MEXT as a social and scientific priority issue (creation of new functional devices and high performance materials to support next-generation industries) to be tackled by using post-K computer and JSPS KAKENHI Grant Numbers JP16H03865. The numerical calculations were carried out using the system B and the system C of the Institute for Solid State Physics at the University of Tokyo, the Oakforest-PACS of the Center for Computational Sciences at University of Tsukuba, and the K computer provided by the RIKEN Advanced Institute for Computational Science through the HPCI System Research project (Project ID: hp190172).
\end{acknowledgments}

\bibliographystyle{jpsj}
\bibliography{refs}

\end{document}